\documentclass[manuscript]{aastex}
\usepackage{color}

\begin{document}

\newcommand{\changeP}[1]{{\textcolor{blue}{#1}}}

\received{}
\accepted{}
\journalid{}{}
\articleid{}{}
\paperid{}
\cpright{AAS}{}
\ccc{}

\slugcomment{Resubmitted \today\ to the {\it Astrophysical Journal}}

\shorttitle{Rotational quenching of water due to He}
\shortauthors{Yang et al.}

\title{Rotational quenching of rotationally-excited H$_2$O in collisions with He}

\author{Benhui Yang\altaffilmark{1}, M. Nagao\altaffilmark{2}, 
        W. Satomi\altaffilmark{2}, M. Kimura\altaffilmark{2,3}, 
        and P. C. Stancil\altaffilmark{1}}
\altaffiltext{1}{Department of Physics and Astronomy and the Center for
     Simulational Physics, The University of Georgia,
         Athens, GA 30602;
yang@physast.uga.edu, stancil@physast.uga.edu}
\altaffiltext{2}{Graduate School of Sciences, Kyushu University, Fukuoka 812-8581, Japan}
\altaffiltext{3}{Deceased}

\begin{abstract}
Theoretical rotational quenching cross sections and rate coefficients of ortho- and para-H$_2$O
due to collisions with He atoms are presented.
The complete angular momentum close-coupling approach as well as the
coupled-states approximation for angular momentum decoupling were applied
to solve the scattering problem for a large range of rotationally-excited states
of water. Results are obtained for 
quenching from initial levels 1$_{1,0}$, 2$_{1,2}$, 2$_{2,1}$, 3$_{0,3}$, 
3$_{1,2}$, 3$_{2,1}$, 
4$_{1,4}$, 3$_{3,0}$, and 4$_{2,3}$ of ortho-H$_2$O and from initial levels 
1$_{1,1}$, 2$_{0,2}$, 2$_{1,1}$, 2$_{2,0}$, 3$_{1,3}$, 3$_{2,2}$, 4$_{0,4}$, 4$_{1,3}$, 
and 3$_{3,1}$ of para-H$_2$O for kinetic energies from 
10$^{-5}$ to 10$^4$ cm$^{-1}$. State-to-state and total deexcitation cross sections 
and rate coefficients for temperatures between 0.1 and 3000 K are reported. 
The present state-to-state rate 
coefficients are found to be in good agreement with previous results 
obtained by Green and coworkers at  
high temperatures, but significant discrepancies are obtained at
lower temperatures likely due to differences in the adopted potential
energy surfaces.
Astrophysical applications of the current rate coefficients
are briefly discussed.

\end{abstract}


\section{INTRODUCTION}
Water is one of the most important molecules in a large variety of astrophysical
environments. As such it is the focus of a Key program for the {\it Herschel
Space Observatory}, Water In Star-forming regions with {\it Herschel} (WISH). 
WISH is designed to probe the physical and chemical 
structures of young stellar objects using water and related 
molecules and to follow the water abundance from collapsing clouds 
to planet-forming disks \citep{dis11}. Water has been detected in
observations with {\it Herschel} in the
massive star-forming region W3 IRS5 \citep{cha10}, in dark regions \citep{cas11},
 and in protoplanetary disks \citep{car08,sal11}, for example.
Water was also detected in earlier observations by the {\it Submillimeter Wave Astronomy Satellite (SWAS)}, the
{\it Infrared Space Observatory (ISO)}, {\it Odin}, and the {\it Spitzer
Space Telescope}. 
For example, {\it SWAS} observed  the
1$_{10}$-1$_{01}$ 556.936 GHz transition of ortho-water, which 
revealed the presence of widespread emission and absorption by water 
vapor around the strong submillimeter continuum source Sagittarius B2 \citep{neu03}.
Detections of thermal water vapor absorption lines were made toward Orion IRc2 
using the Short Wavelength Spectrometer (SWS) on board {\it ISO} \citep{wri00}.
\citet{fra08} presented detections of the 1$_{10}$-1$_{01}$  line
in 18 molecular outflows with {\it SWAS}, while
{\it Spitzer} presented the first clear signature of water vapor on a hot, 
gas planet outside our solar system, HD 189733b  \citep{gri08}.
A large number of H$_2$O lines were also observed with {\it Spitzer} from the
outflow of NGC 2071 \citep{mel08}
and studied with shock models by \citet{flo10}.

The significance of water arises primary due to its importance as a coolant
\citep[e.g.,][]{dot97}, but it is also responsible for observed maser action.
Water observations serve as an important tool for studying 
long-sought details of the planet formation process \citep{ehr07}.
It is the most abundant molecule frozen on 
grain surfaces and is found to change the optical properties of grains and to 
aid in the coagulation process that ultimately produces planets \citep{whi01}. 

To interpret observations of water emission lines, accurate line frequencies and
oscillator strengths are needed, but also key are the availability and 
 accuracy of collisional rate coefficients.
Collisional state-to-state rate coefficients are of great importance to 
describe  energy exchange processes responsible for
thermal balance and line formation, 
particularly in low-density gas. In such situations, the molecular 
level populations may be out of equilibrium or driven to non-local
thermodynamic equilibrium (NLTE) via external energy sources.

Most studies of collisional excitation begin with the consideration of
He impact as the resulting collisional complex is weakly bound and of
lower dimensionality. As such,
the H$_2$O-He collisional system has been investigated in a number of
experimental \citep{bic75, sla79, bru02, cap05, aqu05,dic10,yan10a,yan10b}
and theoretical \citep{gre80,pal88a,pal88b,gre89,gre91,mal92,gre93,tao96,
kuk93,cal03,hod02,pat02,yang07,mak08,dag10} studies. 
For numerical astrophysical models, 
quantitative determinations of state-to-state rate
coefficients for collisions of water are crucial.

As direct measurements of collisional rate coefficients are generally difficult, 
numerical models often rely on calculations. However, other types
are measurements can be performed to give some insight to the
process and the reliability of calculations. For example,
pressure broadening of five transitions of water due to helium and 
molecular hydrogen was
investigated for temperatures relevant to the cold interstellar medium
by \citet{dic10}.
In the case of He, they found a significant temperature dependence
which is in agreement with results deduced from previous 
theoretical inelastic studies. The situation for H$_2$ is
less satisfactory and requires further study both experimentally
and theoretically \citep[see, for example,][]{tak11}.

Recently, state-to-state differential cross section (DCS) measurements
have been performed for rotational excitation of water by He and H$_2$
\citep{yan10a,yan10b}. The experimental data were compared to DCS
calculations
obtained with the close-coupling method using the potential energy surfaces (PESs)  
for H$_2$O-He \citep{hod02} and H$_2$O-H$_2$ \citep{fau05,val08}.  
Very good agreement 
was found for most transitions providing some confidence
in the reliability of the adopted PESs.

Four PESs \citep{cal03,hod02,pat02,mak08} 
have been developed recently for the H$_2$O-He complex.  
The potentials of \citet{hod02} and \citet{pat02} were  
constructed using symmetry-adapted perturbation theory (SAPT), 
referred to as SAPT-H and SAPT-P, respectively. 
\citet{cal03} reported their potential based on valence bond (VB) calculations.  
Detailed comparison of the dynamic performance on the first three potentials, SAPT-H, SAPT-P,
and VB,  was performed by
\citet{yang07}, who concluded that the SAPT-P potential \citep{pat02} was likely
the most reliable.
The PES of \citet{mak08} is expected to be of a similar quality, but has not been
adopted in scattering calculations to the best of our knowledge.

In this work, we extend the previous rotational quenching calculations of
\citet{yang07} to higher levels of rotational excitation primarily utilizing
the  SAPT-P \citep{pat02} PES
for both para- and ortho-water.
H$_2$O-He rate coefficients are presented for a large range of temperature which will
be applicable to modeling a wide variety of astrophysical and atmospheric
environments.

\section{THEORETICAL METHOD}

The theory for scattering of a asymmetric top, such as H$_2$O, with 
a spherical atom can be found in \citet{gar76} and \citet{gre76}.
The center of the coordinate system is fixed at the center of mass of H$_2$O. The water molecule
is located in the $x-z$ plane, with oxygen on the positive $z$-axis. 
H$_2$O is take to be rigid with the O-H bonds and
H-O-H angle fixed at their equilibrium positions. The interaction then
depends only on the position of the He atom which is described in the usual
polar coordinates: $R$, $\theta$, and $\phi$. For the scattering calculations, it is convenient
to expand the angle dependence of the potential $V$ in spherical 
harmonics $Y_{\lambda\mu}(\theta,\phi)$ as:
\begin{equation}
  V(R,\theta, \phi)=\sum_{\lambda\mu}v_{\lambda\mu}(R)(1+\delta_{\mu 0})^{-1}
   [Y_{\lambda\mu}(\theta,\phi)+(-1)^{\mu}Y_{\lambda, -\mu}(\theta, \phi)].
\end{equation}
 Owing to the $C_{2v}$ symmetry of water, only even values of $\mu$ enter the expansion.
The rotational levels of H$_2$O are labeled by  
$j_{k_{-1},k_{+1}}$, where $k_{-1}$ and $k_{+1}$ are the $k$ quantum numbers
in the prolate and oblate limits.
 Rotational wave functions were obtained by diagonalizing the rigid-rotor Hamiltonian
using rotational constants, $A=27.881$ cm$^{-1}$, $B=14. 522$ cm$^{-1}$,  and $C=9.278$ cm$^{-1}$, 
with rotational energy levels taken from  \citet{kyr81} and \citet{lan99}. 
Energy levels of the first 12 rotational states of ortho- and para-H$_2$O 
are given in Table~1.

The calculations presented in this paper were performed by applying the 
close-coupling (CC) approach and the coupled-states (CS) approximation
\citep[see, for example,][]{flo07} and using the SAPT-P PES, except where noted.
The CC method was used to calculate rotational quenching cross sections 
for center-of-mass kinetic energies from 10$^{-5}$ to 450 cm$^{-1}$, while the
CS approximation was used 
from 500 to 10,000 cm$^{-1}$.
All the CC and CS calculations  were performed using 
the nonreactive scattering code MOLSCAT \citep{molscat}. 
In the quantum scattering calculations, 
the coupled-channel equations were integrated using the modified
log-derivative Airy propagator of \citet{ale87} with a variable step-size.
The propagation was carried out to a maximum intermolecular separation of
$R=100$  \AA.  At each energy, a sufficient
number of total angular momentum partial waves was included to ensure
convergence of the cross sections. The maximum value of the total angular
momentum quantum number $J$ 
employed in the calculations was 140.  
$\vec{J}=\vec{l}+\vec{j}$ and $l$ is the orbital angular momentum
of the complex.
All calculations include 5 to  10 energetically closed channels to ensure 
converged cross sections.
As a consequence of the nuclear spins of the two hydrogen atoms in H$_2$O,
water exists in two forms: para-H$_2$O and ortho-H$_2$O. The ortho- and 
para-levels do not
interconvert in nonreactive collisions and are therefore 
treated separately in the current inelastic calculations.

The cross sections  were
thermally averaged over a Maxwellian kinetic energy distribution to yield state-to-state
rate coefficients  from specific initial rotational states $j_{k_{-1},k_{+1}}$
as functions of the temperature $T$,
\begin{equation}
k_{j_{k_{-1}, k_{+1}}\rightarrow j^{\prime}_{k^{\prime}_{-1},k^{\prime}_{+1}}}(T) 
= \left (\frac{8}{\pi m \beta} \right )^{1/2}\beta^2\int^{\infty}_0 E_k
\sigma_{j_{k_{-1}, k_{+1}}\rightarrow j^{\prime}_{k^{\prime}_{-1},k^{\prime}_{+1}}}(E_k) 
\exp(-\beta E_k)dE_k
\label{eq2}
\end{equation}
where $\sigma_{j_{k_{-1}, k_{+1}}\rightarrow j^{\prime}_{k^{\prime}_{-1},k^{\prime}_{+1}}}(E_k)$ 
is the rotational transition cross section with
$j_{k_{-1},k_{+1}}$ and $ j^{\prime}_{k^{\prime}_{-1},k^{\prime}_{+1}}$ being, respectively, 
the initial and final rotational states of
H$_2$O, $m$ is the reduced mass of the He-H$_2$O complex, $E_k$ the kinetic energy, and
$\beta=(k_BT)^{-1}$, where $k_B$ is Boltzmann's constant.

\section{RESULTS AND DISCUSSION}

\subsection{State-to-state and total deexcitation cross sections}

\begin{figure}
\epsscale{0.9}
\plotone{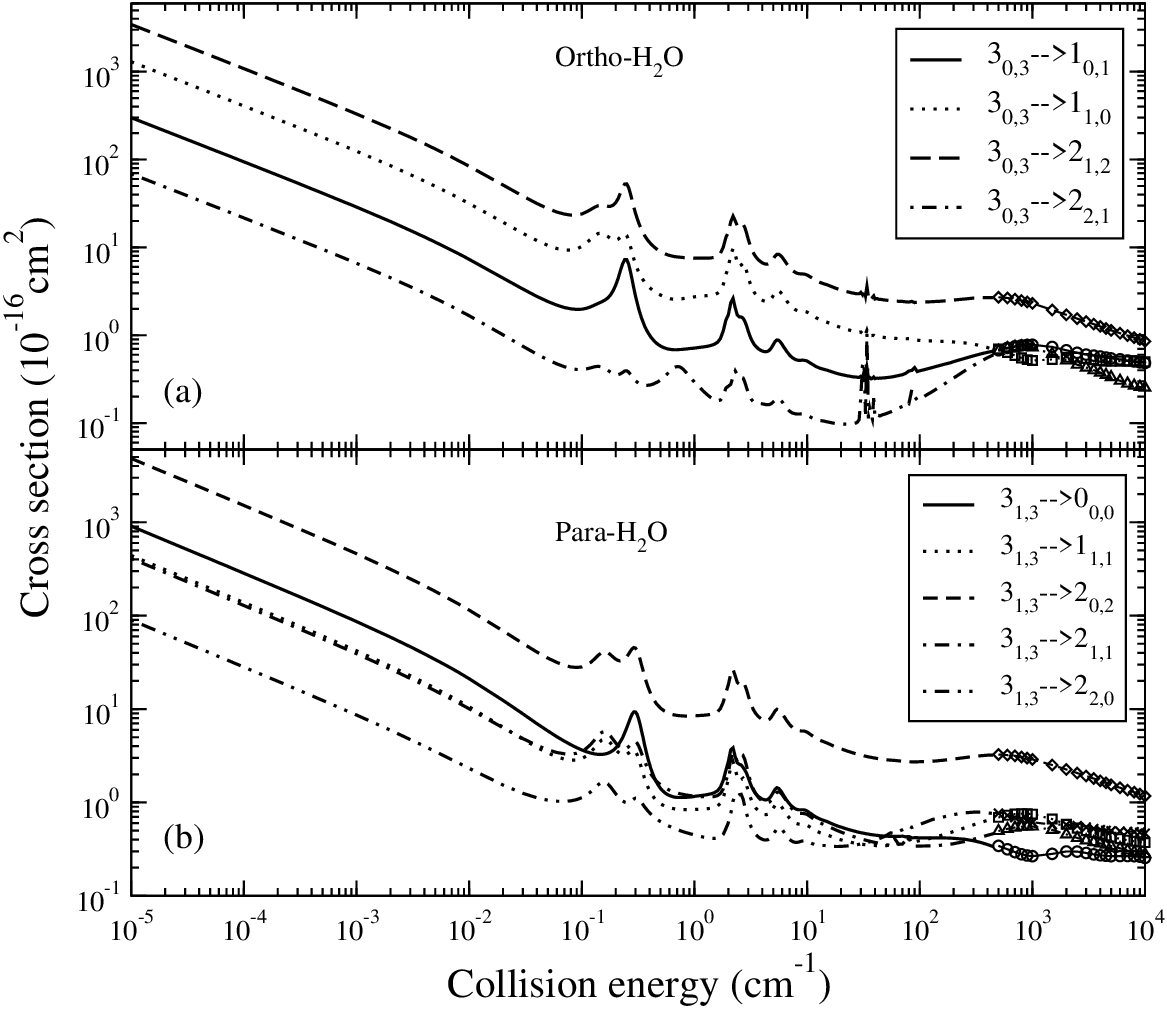}
\caption{
State-to-state deexcitation cross sections of H$_2$O in collisions with He as a function 
of kinetic energy obtained with the CC method (lines) and the CC 
approximation (symbols).  (a) 3$_{0,3}$ of ortho-H$_2$O,  (b) 3$_{1,3}$ of para-H$_2$O. 
}
\label{fig1}
\end{figure}

State-to-state deexcitation cross sections were computed for initial rotational
levels 1$_{1,0}$, 2$_{1,2}$, 2$_{2,1}$, 3$_{0,3}$, 3$_{1,2}$, 3$_{2,1}$, 
4$_{1,4}$, 3$_{3,0}$, and 4$_{2,3}$ of ortho-H$_2$O and initial levels 
1$_{1,1}$, 2$_{0,2}$, 2$_{1,1}$, 2$_{2,0}$, 3$_{1,3}$, 3$_{2,2}$, 4$_{0,4}$, 4$_{1,3}$,
and 3$_{3,1}$ of para-H$_2$O, extending the results from our earlier
work \citep{yang07} which was limited to $j\leq 2$.
As examples, the state-to-state quenching cross sections from the initial level 3$_{0,3}$ 
of ortho-H$_2$O and 3$_{1,3}$ of para-H$_2$O are presented 
in Figs.~\ref{fig1}(a) and \ref{fig1}(b), respectively.\footnote{All state-to-state 
cross sections and rate coefficients for quenching
are available on the UGA Molecular Opacity Project website
(www.physast.uga.edu/ugamop/). The rate coefficients are also available in
the format of the Leiden Atomic and Molecular Database \citep[LAMDA,][]{sch05}
on our website.}
The agreement between CC and CS calculations, as shown in Fig.~\ref{fig1}, is found to be
excellent. We conclude that the CS approximation, which is computationally
efficient, is reliable for the current collision system at high energies.

From Fig.~\ref{fig1} it can be seen that the cross sections display resonances 
in the intermediate energy region from 0.1 to 100 cm$^{-1}$ due to the influence of the 
attractive part of the interaction potential. The energy location and magnitude
of the resonances are sensitive to the details of the PES as shown in
\citet{yang07}. Importantly, for astrophysical applications, 
the properties of the resonances influence the quenching rate coefficients
at low temperatures as discussed below.
The 3$_{0,3}\rightarrow$ 2$_{1,2}$ transition is seen to dominant the 
quenching of the 3$_{0,3}$ level and likewise for para-H$_2$O, the 
dominant quenching transition for the 3$_{1,3}$ is 
3$_{1,3}\rightarrow$ 2$_{0,2}$. Both transitions are seen to obey the
propensity rule $|\Delta j| = |\Delta k_{+1}|=|\Delta k_{-1}| = 1$. 

\begin{figure}
\epsscale{1.05}
\plottwo{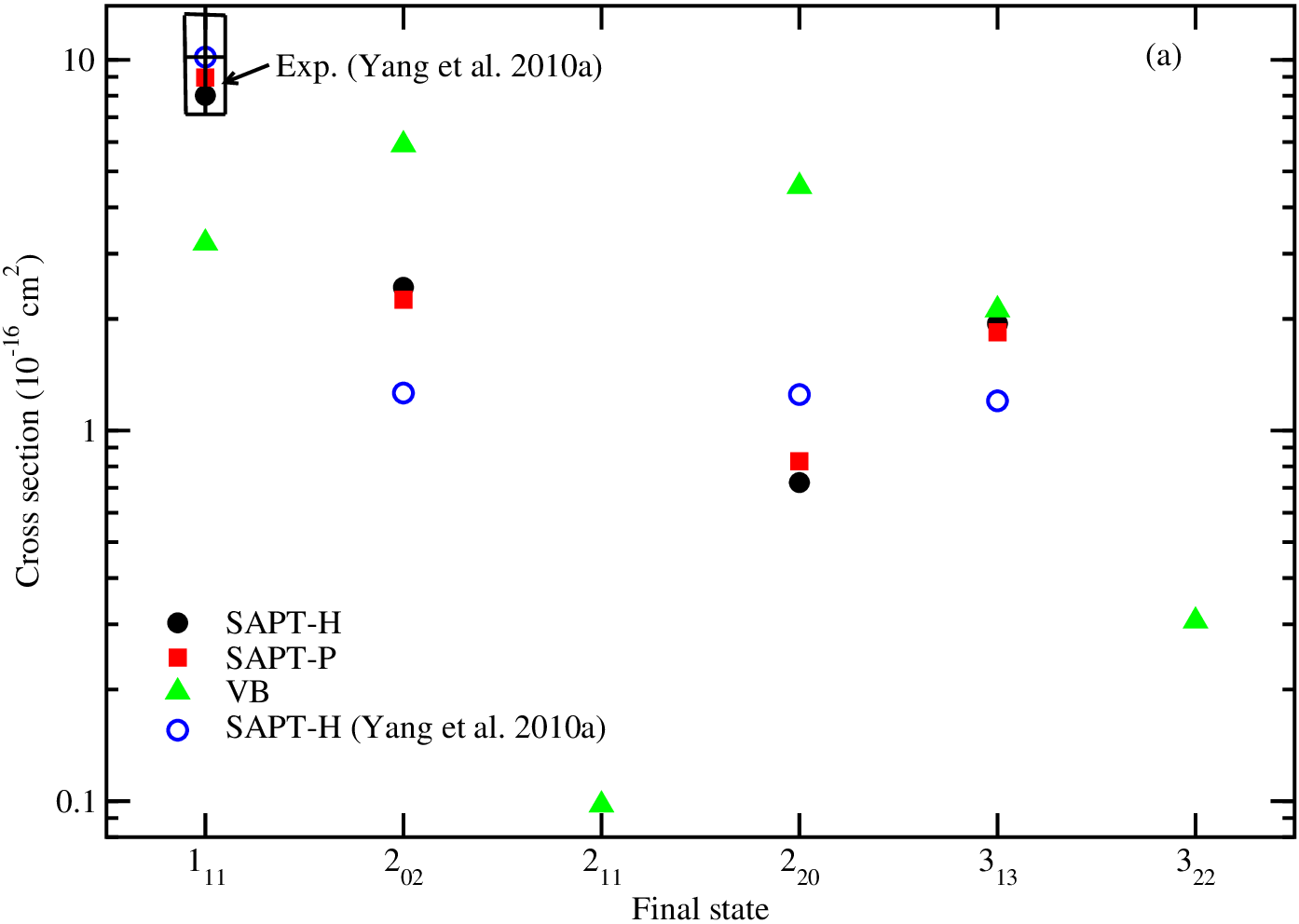}{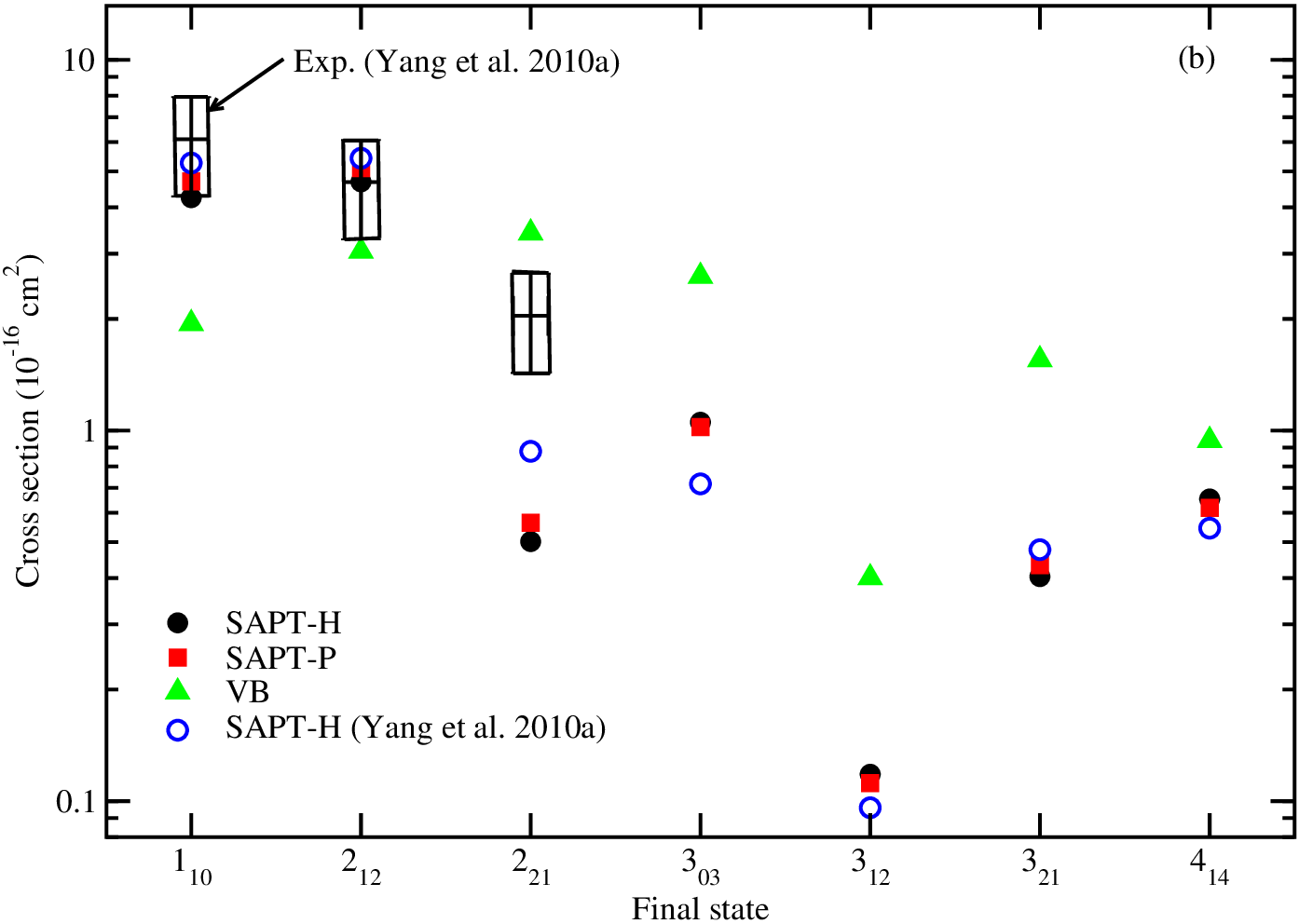}
\caption{Comparison of state-to-state H$_2$O-He excitation cross sections at
429 cm$^{-1}$ from the initial states (a) $0_{00}$ and (b) $1_{01}$. Open
circle, CC calculations of \citet{yan10a}; filled symbols, current CC calculations;
boxes, experiment of \citet{yan10a}. Note that in (a) the cross section magnitudes
from other calculations to the $2_{11}$ and $3_{22}$ are smaller than the scale of the plot.
}
\label{fig2}
\end{figure}

\begin{figure}
\epsscale{0.8}
\plotone{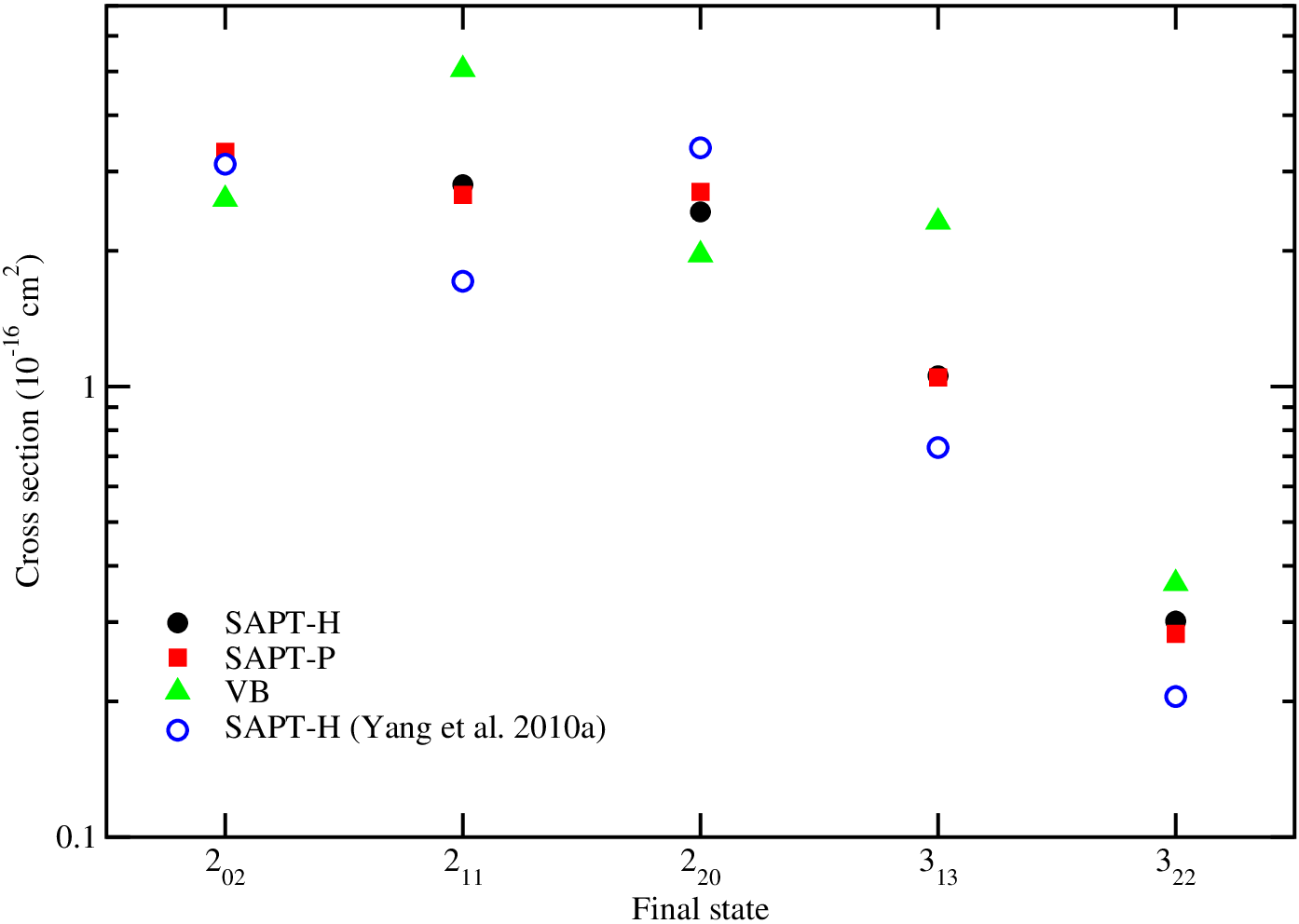}
\caption{Same as Fig. 2, but for the initial state $1_{11}$. No
experimental integral cross sections are available for this state.
}
\label{fig3}
\end{figure}

As a test of the accuracy of our results, we can compare 
to the recent measurements and calculations of \citet{yan10a}. In
their work, relative DCS measurements of H$_2$O state-to-state
excitation due to He impact were performed at a kinetic energy of
429 cm$^{-1}$ for 12 different transitions. Differential and integral
cross sections were also computed using MOLSCAT, but with the SAPT-H PES. They found good agreement
between theory and measurement
in all cases except for the $0_{00}\rightarrow 2_{11}$ and
$0_{00} \rightarrow 3_{22}$ transitions \citep{yan10a}. As these transitions are weak (see
Fig.~\ref{fig2}), the authors argued that the discrepancy may be due to contamination
of their beam of ground state molecules ($0_{00}$) with excited $1_{11}$ water. For four dominant transitions, the relative
DCSs were used to obtain relative integral cross
sections which the authors normalized to the largest absolute cross
section from their calculations,
$0_{00}\rightarrow 1_{11}$.  These are reproduced in Fig.~\ref{fig2} adopting the
quoted experimental cross section uncertainty of  $\pm30\%$, while the
uncertainty in the beam energy was $\pm20\%$.  Also, shown in Fig.~\ref{fig2}
and Fig.~\ref{fig3} are
the current calculations using the SAPT-P PES, as well as new
results using the SAPT-H and VB surfaces, all performed at a kinetic 
energy of 429 cm$^{-1}$. As noted in \citet{yang07},
the VB surface gives cross sections that typically differ significantly 
with results obtained on other surfaces and with available experimental
data. This is further confirmed in the present work and with comparison
to the \citet{yan10a} measurements. For all considered
transitions, the current calculations using the SAPT-H and SAPT-P PESs result
in similar cross sections. This is consistent with the study performed in
\citet{yang07} which had difficulty in selecting a preferred PES between the two.
Significant differences between the current calculations
and those of \citet{yan10a} are noted, even though both sets of 
calculations adopted the same surface (SAPT-H)
and the same scattering code. For the current calculations, all parameters
were carefully checked to ensure convergence and we can offer no explanation
for the difference. Finally, the agreement with
experiment, regardless of normalization selection, is seen to be excellent 
for all the calculations for all adopted surfaces (except, of course, for
the VB surface). The one exception occurs for the transition $1_{01}\rightarrow 2_{21}$.
\citet{yan10a} do not comment on this  discrepancy and we can only
speculate that their beam of $1_{01}$ molecules may have been contaminated
by other low-lying excited states which may have influenced the cross section
measurement given its relatively small magnitude. We conclude that the current
calculations using the SAPT-P surface gives results of sufficient reliability
for astrophysical needs.

\begin{figure}
\epsscale{0.9}
\plotone{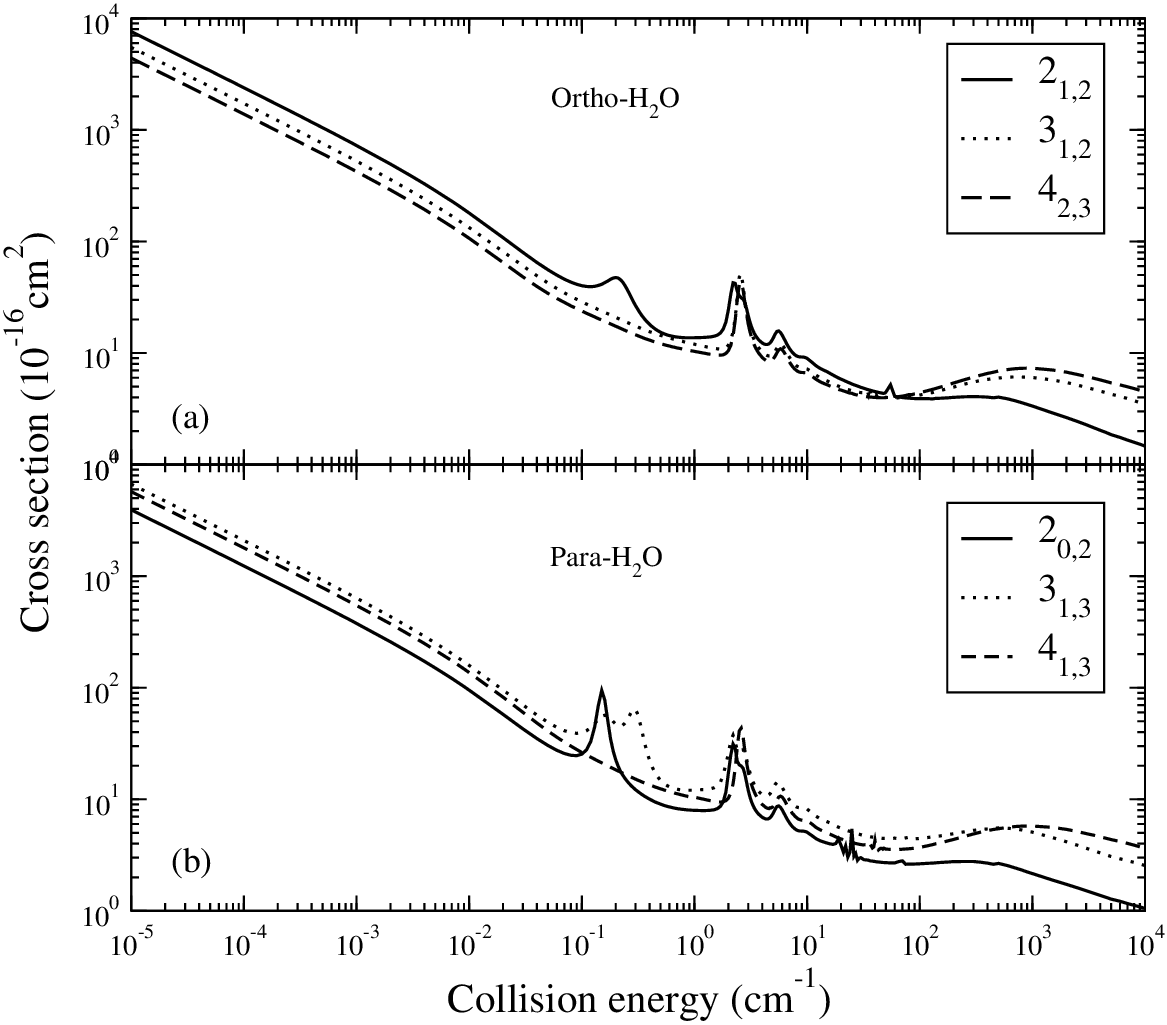}
\caption{
Total deexcitation cross sections from states: (a) 2$_{1,2}$, 3$_{1,2}$, and 4$_{2,3}$ of
ortho-H$_2$O,  and (b) 2$_{0,2}$, 3$_{1,3}$, and 4$_{1,3}$
of para-H$_2$O in collisions with He as a function of kinetic energy.
}
\label{fig4}
\end{figure}

State-to-state cross sections from each initial state are summed over all
final states to obtain the total deexcitation cross section. 
In Fig.~\ref{fig4}, we compare the total deexcitation cross sections from
select initial levels 
2$_{1,2}$, 3$_{1,2}$, and 4$_{2,3}$ of ortho-H$_2$O and
2$_{0,2}$, 3$_{1,3}$, and 4$_{1,3}$ of para-H$_2$O, while cross
sections from some of the lower $j$ levels are given in
\citet{yang07}.   
Generally, the total deexcitation cross sections from different initial levels 
have similar behavior and are of similar magnitude over the
considered energy range. Each of the total cross sections 
exhibit the behavior predicted by Wigner's threshold law
\citep{wig48} at ultra-low collision energies, where only $s$-wave scattering
contributes and the cross sections vary inversely with the relative velocity. 
In the intermediate energy region, between 0.1 and 
100 cm$^{-1}$, the cross sections 
display scattering resonances. As a consequence of the diverse behavior
of the resonances, no trends are evident except at the highest energies 
where the cross sections increase with rotational excitation.

\subsection{State-to-state and total deexcitation rate coefficients}

\begin{figure}
\epsscale{0.9}
\plotone{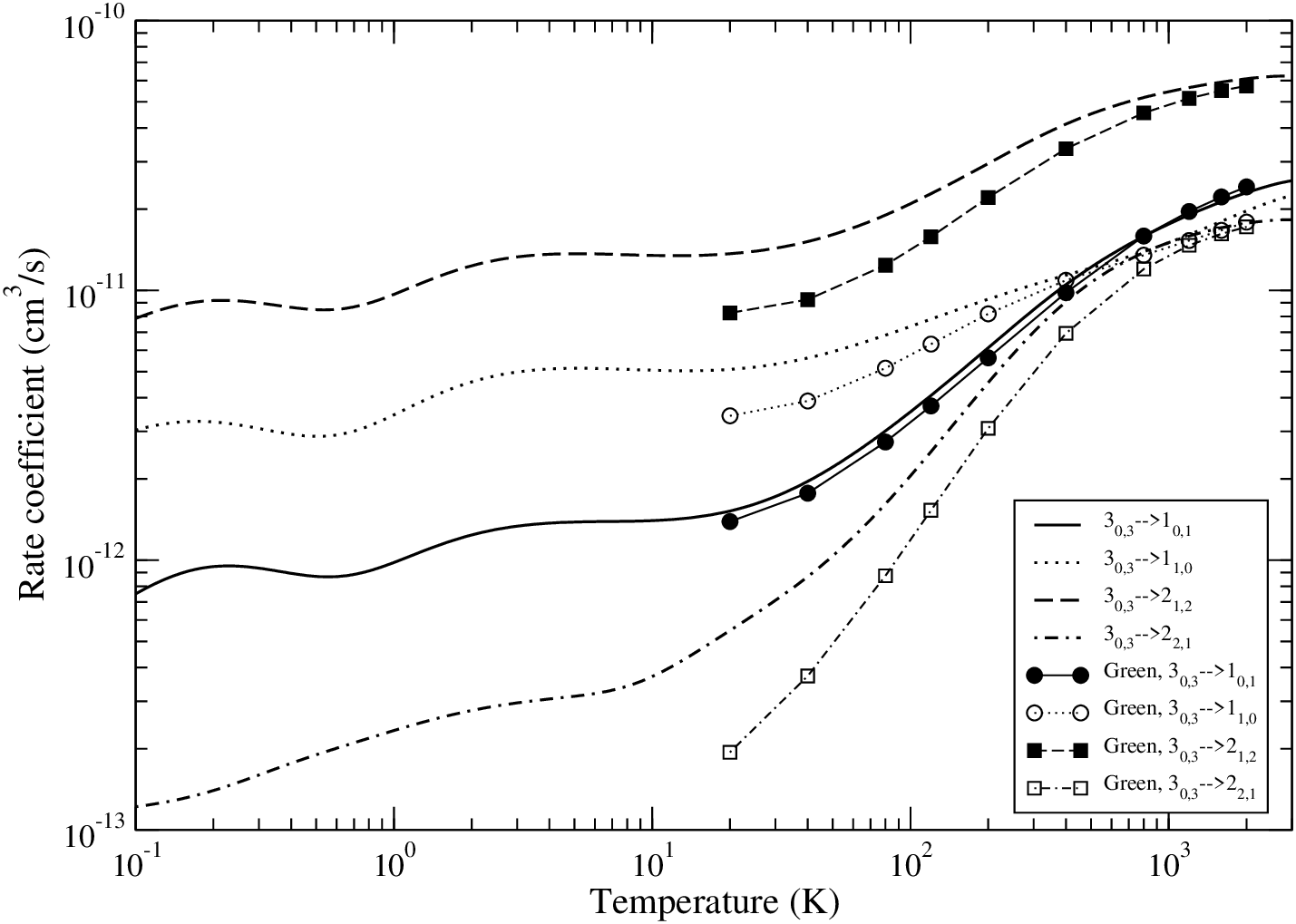}
\caption{ 
The state-to-state deexcitation rate coefficients from initial state 3$_{0,3}$ of ortho-H$_2$O by
collisions with He atoms as a function of temperature. Current results (lines) and 
results from \citet{gre93} (symbols with lines). 
}
\label{fig5}
\end{figure}

\begin{figure}
\epsscale{0.9}
\plotone{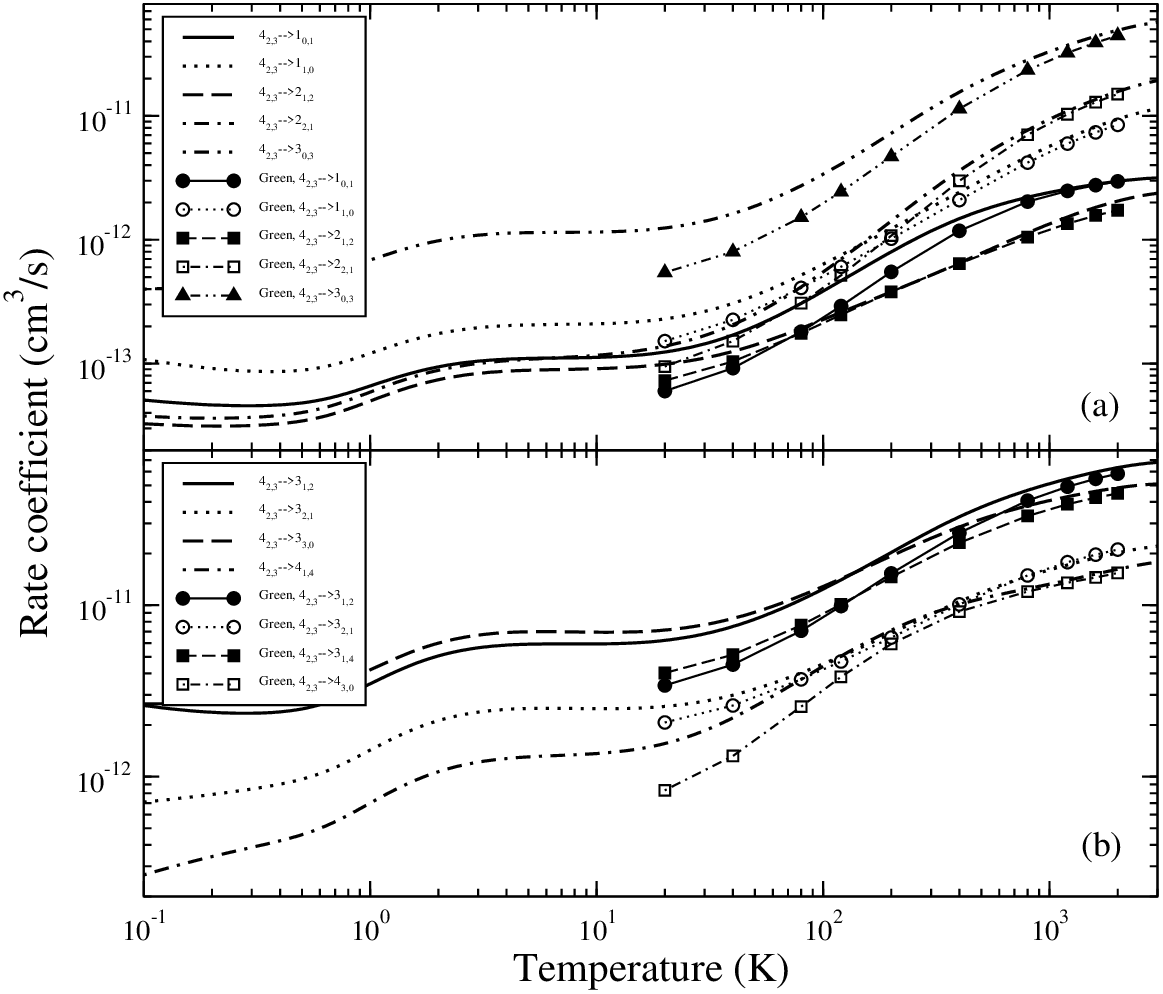}
\caption{
Same as Fig.~\ref{fig5}, but for ortho-H$_2$O initial state 4$_{2,3}$.
}
\label{fig6}
\end{figure}

\begin{figure}
\epsscale{0.9}
\plotone{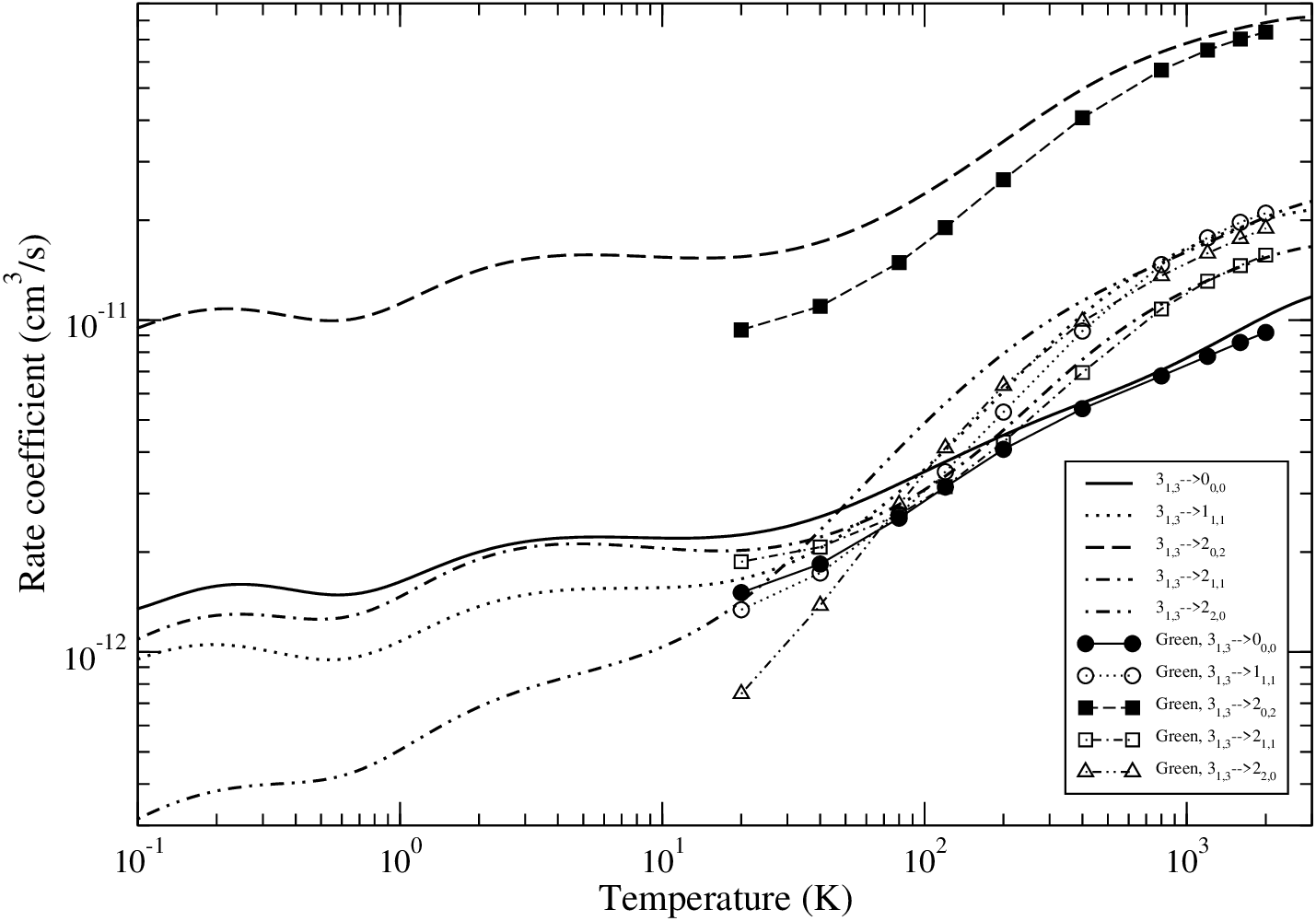}
\caption{
Same as Fig.~\ref{fig5}, but for para-H$_2$O initial state 3$_{1,3}$.
}
\label{fig7}
\end{figure}

\begin{figure}
\epsscale{0.9}
\plotone{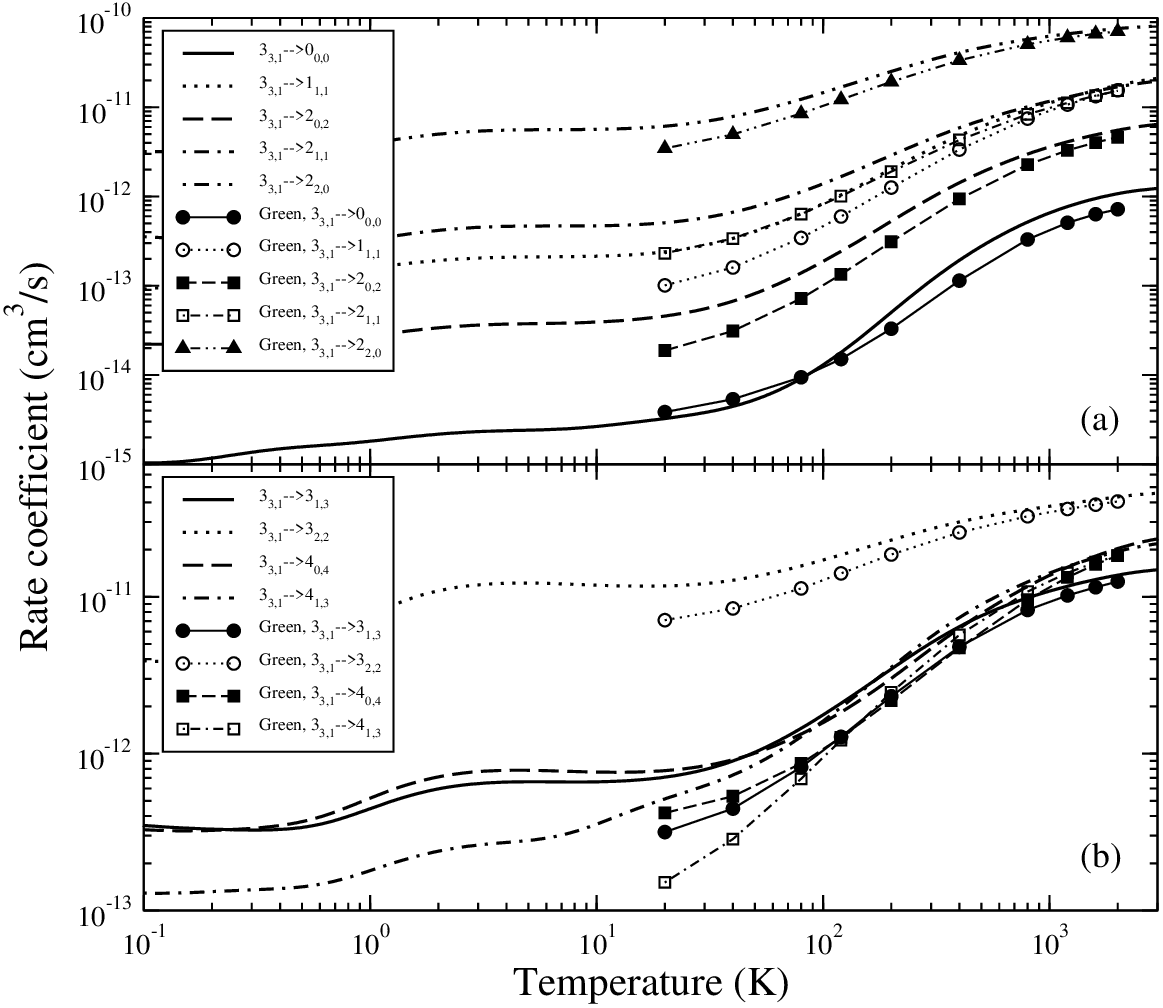}
\caption{
Same as Fig.~\ref{fig5}, but for para-H$_2$O initial state 3$_{3,1}$.
}
\label{fig8}
\end{figure}

Deexcitation rate coefficients for temperatures ranging from 0.1 
to 3000 K are shown in Figs.~\ref{fig5}-\ref{fig8} for some select levels
of ortho-H$_2$O and 
para-H$_2$O. The rate coefficients are given below 3~K 
to illustrate their behavior as they approach the zero-temperature
limit.
Unfortunately, we are unaware of any
experimental rate coefficient data for rotational transitions of H$_2$O due to 
collisions with He. Therefore, we can only compare our computations with the 
theoretical results of \citet{gre93}, which were calculated with the 
potential of \citet{mal92}.
The \citet{gre93} calculations were  carried out using MOLSCAT, 
with the modified log-derivative Airy propagator, and using the
CC method for a total energy up to 470 cm$^{-1}$. For para-H$_2$O (ortho-H$_2$) as
many as 17 (16) rotational levels were included in their basis. For higher total energies,
\citet{gre93} used the CS approximation up to  5000 cm$^{-1}$. The primary difference
between the current calculations and those of \citet{gre93} is the adopted PES.

Figs.~\ref{fig5} and \ref{fig6}, which display state-to-state quenching rate coefficients  
for ortho-H$_2$O from initial levels 3$_{0,3}$ and 4$_{2,3}$, respectively,  
 show that between 0.1 and 50~K,
the rate coefficients exhibit an oscillatory temperature dependence due to
the presence of resonances.
For temperatures above $\sim$50~K, the rate coefficients generally
increase with increasing temperature.
Comparison with the rate coefficients of \citet{gre93} shows
generally good agreement at temperatures above $\sim$200 K. 
The results of \citet{gre93} are smaller than the 
present rate coefficients for lower
temperatures with the discrepancy increasing with decreasing temperature.
The differences are primarily related to the adopted PESs since all other
aspects of the two sets of calculations appear to be similar, though 
different choices of convergence parameters are possible.
Further, the bumps in the present rate coefficients are 
absent from the results of Green et al., likely due to resonances not 
produced for scattering on the \citet{mal92} PES.

Considering scattering of para-H$_2$O by He, the trends noted for ortho-H$_2$O are
also evident in the state-to-state deexcitation rate coefficients as shown 
in Figs.~\ref{fig7} and \ref{fig8} for some example initial states.
Oscillations due to resonances and
a general increase in the rate coefficients above 50~K are similar.
Comparisons to the work of Green et al. generally show good agreement, but again only for temperatures above $\sim$200 K.

\begin{figure}
\epsscale{0.9}
\plotone{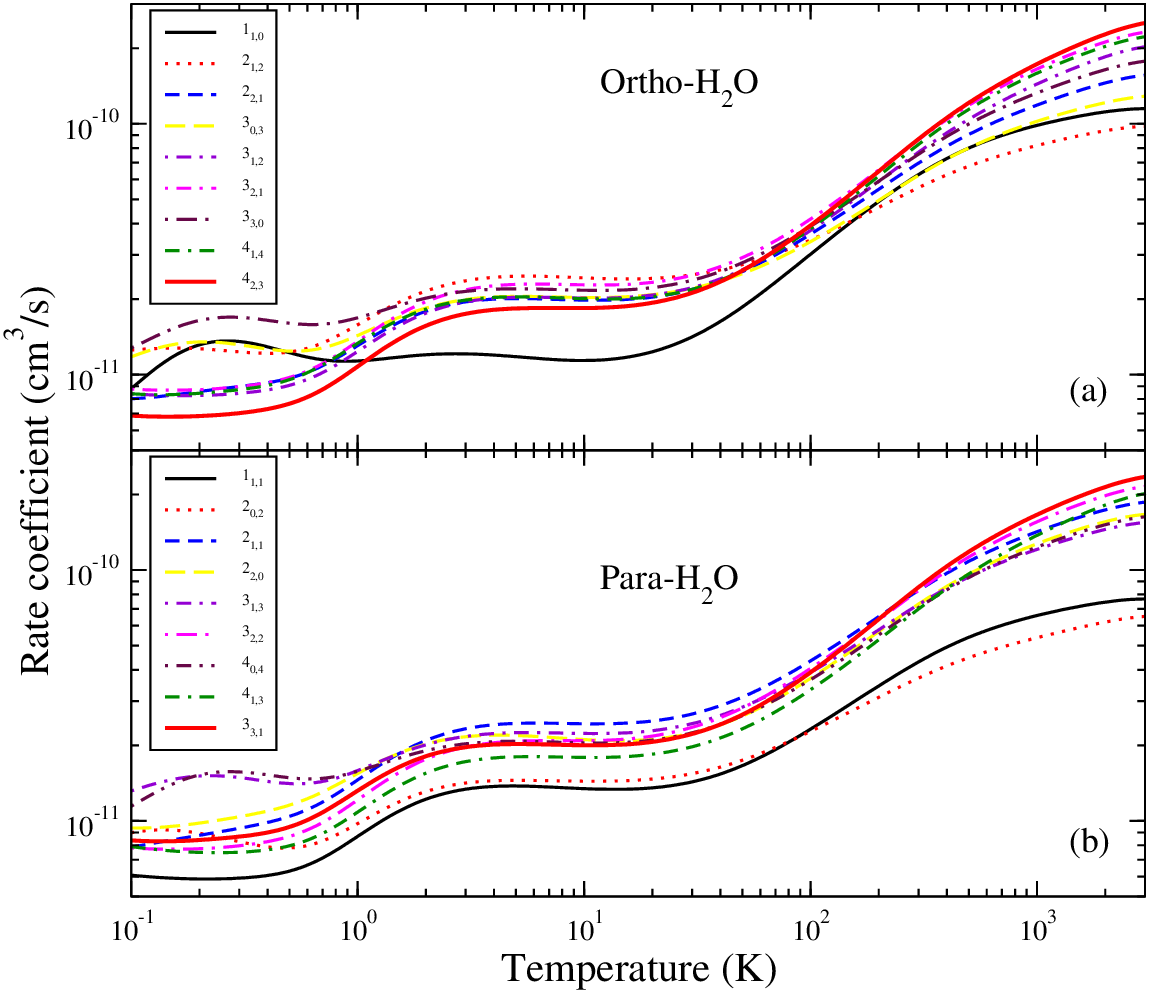}
\caption{
The total H$_2$O-He excitation rate coefficients as a function of temperature.
(a) Ortho-H$_2$O,  (b) para-H$_2$O. 
}
\label{fig9}
\end{figure}

The total deexcitation rate coefficients from all considered initial states
\citep[including those given in][]{yang07} are shown in Fig.~\ref{fig9}. They
exhibit similar oscillatory behavior, but the varying oscillation dependencies reflect
different resonance energies. 
For temperatures above $\sim$50~K, the total deexcitation rate coefficients
smoothly increase with increasing temperature, a trend which may be useful in
scaling arguments to estimate rate coefficients for more highly excited states not
explicitly computed.

\section{ASTROPHYSICAL APPLICATIONS}

As discussed in the Introduction, water has been observed in emission and absorption 
in a variety of astronomical environments.  For example, in the past decade there has been increasing
interest in the properties of protoplanetary disks (PPDs) and considerable effort has
gone into the modeling of PPDs in general and the spectral features of water, in particular
\citep{gor04,mei08,mei09,gla09,sal11}. The observation of water in PPDs has become
very common \citep{car08,car11,sal11} with interest focused on the inner-planet forming
region. \citet{dul10} point out that water is one of dominant coolants in the inner regions.

Because the critical densities of water are typically greater than 10$^8$ cm$^{-3}$, models
which assume a Boltzmann level population distribution, or local thermodynamic
equilibrium (LTE), are generally not valid. Full NLTE models are necessary
to predict the rotational and vibrational spectrum of water in most emitting regions
of PPDs, above the plane of the disk, as demonstrated by \citet{mei09} and in a variety of other environments.
NLTE models require knowledge of
state-to-state rotational (and vibrational) rate
coefficients for the kinetic modeling of the internal level populations.
Important colliders may include H$_2$, H, He, H$^+$, and e$^-$, and
for comets H$_2$O \citep{ben04}. A review of the status of collisional rate
coefficients for water can be found in \citet{tak11}. Because of the complexity
in considering H$_2$ collisions, there is a long history of initial collisional excitation
calculations being performed with He collisions. The He rate coefficients are then
often mass-scaled to emulate para-H$_2$ collisions. We certainly {\it do not} advocate
such a procedure here and it has been pointed out by \citet{dub09} that mass-scaling
does not result in reliable H$_2$O-H$_2$ rate coefficients. As a consequence, a
considerable amount of effort has been expended to directly compute rate coefficients for 
rotational excitation of water
by H$_2$ \citep{phi95,phi96,dub02,fau06,fau07,dub09,dan10,dan11}\footnote{In particular for H$_2$O-H$_2$, the latest 
state-to-state rate coefficients of \citet{dan11},  with a package to 
calculate effective and thermal rates, can be found on BASECOL.}
, while the rate coefficients
for vibrational excitation are limited to the quasi-classical calculations of \citet{fau05},
with extrapolations proposed by \citet{fau08}. We are unaware of any collisional
data for H$^+$ impact, we have performed some preliminary calculations
for H collisions (Yang \& Stancil, in preparation), while \citet{fau08} have compiled the available
electron collision data. 
However, except for the set of
calculations done for  temperatures between 20 and 2000 K by \citet{gre93}
on the earlier, and
presumably less accurate PES of \citet{mal92}, our previous work \citep{yang07}, and
the small set of calculations in support of the experiments of \citet{yan10a,yan10b},
no other computations for rate coefficients due to He impact have been published.
Therefore, the current rate coefficient calculations,
which extend from 10$^{-1}$~K to 3000~K, are the most comprehensive
to date for He and can be utilized in a variety of applications augmenting the
datasets developed for H$_2$ and electrons 
\citep[see also, the LAMDA website and BASECOL website,][]{sch05,dub06}.
We note that in NLTE models of water in PPDs, \citet{mei09} mass-scaled the
available H$_2$O-H$_2$ rate coefficients to estimate values for H$_2$O-He.

\section{CONCLUSION}

Cross sections and rate coefficients for rotational quenching of ortho- and para-H$_2$O 
due to He collisions have been studied using the close-coupling method and the coupled-states 
approximation on the ab initio potential energy surface of \citet{pat02}.  
State-to-state and total deexcitation cross sections and rate coefficients
for the first 9 initial excited rotational levels of para- and ortho-H$_2$O 
are obtained over a wide temperature range from 0.1 to 3000~K 
and are available in tables formatted
for astrophysical applications.  Numerous orbiting resonances result in
undulations in the temperature dependence of the rate coefficients.
For temperatures less than $\sim$200~K,
the current state-to-state rotational rate coefficients
are found to depart from the results of \citet{gre93}
obtained with the earlier PES of \citet{mal92}.
Discrepancies are primarily related to differences in the adopted PESs as
all other aspects of the current calculations are similar to those of
\citet{gre93}.

\acknowledgements
We dedicate this manuscript to the memory of Professor Mineo Kimura
who passed away during the completion of the work presented here.
B.H.Y. and P.C.S. acknowledge support from NASA under Grants No. NNX07AP12G
and NNX12AF42G.

{}

\begin{table}
\begin{center}
\caption{Rotational energy levels (cm$^{-1}$) of ortho- and para-H$_2$O \label{tbl-1}}
\begin{tabular}{c c c c c c c}
\tableline \tableline
\multicolumn {2}{c}{ ortho-H$_2$O} & &  & & \multicolumn {2}{c}{ para-H$_2$O}  \\ [0.5ex]
\tableline
1$_{0,1}$ & 23.7943  & & &   &   0$_{0,0}$  &  0.0000 \\   [1ex]
1$_{1,0}$ & 42.3717  & & &   &   1$_{1,1}$  &  37.1371 \\   [1ex]
2$_{1,2}$ & 79.4963  & & &   &   2$_{0,2}$  &  70.0907 \\   [1ex]
2$_{2,1}$ & 134.9018 & & &   &   2$_{1,1}$  &  95.1757 \\   [1ex]
3$_{0,3}$ & 136.7617 & & &   &   2$_{2,0}$  &  136.1641 \\   [1ex]
3$_{1,2}$ & 173.3656 & & &   &   3$_{1,3}$  &  142.2783 \\   [1ex]
3$_{2,1}$ & 212.1561 & & &   &   3$_{2,2}$  &  206.3013 \\   [1ex]
4$_{1,4}$ & 224.8383 & & &   &   4$_{0,4}$  &  222.0529 \\   [1ex]
3$_{3,0}$ & 285.4192 & & &   &   4$_{1,3}$  &  275.4971 \\   [1ex]
4$_{2,3}$ & 300.3621 & & &   &   3$_{3,1}$  &  285.2200 \\   [1ex]
5$_{0,5}$ & 325.3483 & & &   &   4$_{2,2}$  &  315.7792 \\   [1ex]
4$_{3,2}$ & 382.5171 & & &   &   5$_{1,5}$  &  326.6256 \\   [1ex]
\tableline
\end{tabular}
\end{center}
\label{tab1}
\end{table}

\newpage

\end{document}